\input{amssymb.sty}
\documentclass[12pt]{article}
\def\Tc{T_{\mbox{\scriptsize c}}}
\def\rhoc{\rho_{\mbox{\scriptsize c}}}

\newcommand{\ssection}[1]{%
   \section[#1]{\raggedright\bf #1}}

\begin{document}
\input{epsf.sty}
\title{\vspace{-0.9in}\large \bf Fluid Critical Points from Simulations: the Bruce-Wilding method and Yang-Yang anomalies\vspace{-0.25in}}

\author{\vspace{-0.1in}\normalsize Young C.\ Kim and Michael E.\ Fisher \\ \vspace{-0.1in} {\normalsize\em Institute for Physical Science and Technology} \\ \vspace{-0.1in} {\normalsize\em University of Maryland, College Park, Maryland 20742}}

\date{\normalsize\today}

\maketitle
\vspace{-0.3in}
\begin{abstract}

A critique is presented of the frequently used Bruce-Wilding (BW) mixed-field scaling method for estimating the critical points of nonsymmetric model fluids from grand canonical simulation data. An explicit, systematic technique for implementing this method is set out thereby revealing clearly a fortunate, close cancelation of contributions from the leading correction-to-scaling and thermal scaling functions that makes the method effective for Ising-type systems but which lacks a general theoretical basis. The BW approach does not allow for pressure mixing in the scaling fields which is essential for representing a Yang-Yang anomaly, namely, the divergence at criticality of the second temperature derivative, $(d^{2}\mu_{\sigma}/dT^{2})$, of the chemical potential $\mu_{\sigma}(T)$ on the phase boundary; but such behavior must be expected in realistic models. We show that allowance for pressure mixing does not alter the leading dependence of the critical temperature estimator, $\Tc(L)$, on the linear size, $L$, of the simulation box: this converges as $L^{-(1+\theta)/\nu}$ when $L\rightarrow\infty$ (where $\nu\simeq 0.6$ and $\theta\simeq 0.5$ are the correlation-length and leading correction critical exponents). On the other hand, the critical density estimator, $\rhoc(L)$, gains a leading variation $\propto L^{-2\beta/\nu}$ that dominates the previously claimed $L^{-(1-\alpha)/\nu}$ term (where $\alpha\simeq 0.1$ and $\beta\simeq 0.3$ are the specific heat and coexistence curve exponents). Numerically, the BW method provides estimates of $\Tc$ consistent with those obtained from recently developed {\em unbiased} techniques that do {\em not} require prior knowledge of the universal order-parameter and energy distribution functions; however, BW estimates of the critical densities, $\rhoc$, prove significantly less reliable.

\end{abstract}

\pagebreak

\ssection{\hspace{-.27in}.\hspace{0.1in}Introduction and Overview}
\indent

The critical behavior of fluids, including the issues of exponents and universality class and the magnitudes of non-universal quantities, such as the location of critical points, has been much studied by Monte Carlo simulations of model systems with the aid of increasingly powerful computers. However, since simulations always deal with finite systems, say of linear dimensions $L$, a precise calculation can do no more than reveal details of the rounding of the system's properties in the critical region: a sharp critical point is attained only in the limit $L\rightarrow\infty$. Thus the analysis of finite-size effects by suitable scaling concepts and subsequent extrapolation is a key ingredient for the success of such investigations.$^{1}$ 

A notable advance in this respect was made in 1992 when Bruce and Wilding$^{2,3}$ (BW) developed a special finite-size {\em mixed-field} scaling technique or ``recipe'' that has been widely used in deriving critical parameter estimates for a variety of more-or-less realistic model fluids.$^{4-20}$ In their approach the joint distribution function, $P_{L}(\rho,u)$, of the (number) density $\rho=N/V$ and the  configurational energy density, $u$, for a finite-size system of volume $V=L^{d}$ and dimensions $L$$\,\times\,$$L$$\,\times\,$$\,\cdots$$\,\times\,$$L$ with periodic boundary conditions is calculated via grand canonical ensemble simulations of the model fluid at temperature $T$ and chemical potential $\mu$. Within scaling theory BW then described this joint distribution of density and energy fluctuations in a finite near-critical fluid by incorporating the mixing of two {\em relevant scaling fields}; in other words, the ordering field $\tilde{h}$ (conjugate to the order parameter) and the thermal field $\tilde{t}$ (conjugate to the critical energy density) were taken as {\em linear combinations} of the ``bare'' physical fields, $t \propto T - \Tc$ and $h \propto \mu - \mu_{\mbox{\scriptsize c}}$.$^{21-23}$ This mixing of $t$ and $h$ into the scaling fields $\tilde{h}$ and $\tilde{t}$ is a crucial manifestation of the {\em lack} of particle-hole {\em symmetry}. An analysis of a $(d$$\,=\,$$2)$-dimensional Lennard-Jones fluid using Monte Carlo simulations confirmed this mixing.$^{3}$ The observed (or calculated) joint distribution was then related to the {\em universal} critical-point distribution for the Ising model, to which universality class ``typical'' fluids are generally believed to belong. It was found$^{3}$ that the {\em limiting} critical behavior of the density distribution assumed a scaling form, as expected; furthermore, the shape of the distribution could be quite well matched to the known universal order-parameter distribution function appropriate to the Ising class. This served to illuminate the underlying basis of the universality shared by simple fluids and Ising ferromagnets. By such a matching of the distribution functions of density and energy to the limiting universal functions --- that were obtained {\em a priori} via computer simulations of simple, symmetric Ising models$^{5,24,25}$ --- Wilding and subsequent collaborators were able to estimate the critical parameters and the mixing coefficients for more general continuum model fluids.

Although the BW method has proved convenient and straightforward and has been successfully applied to estimating the critical parameters in various models, it suffers from certain weaknesses. One of its weakest points is that various crucial features of the critical behavior of the model under the investigation must be known {\em a priori}. In particular, the method requires precise advance knowledge of the fixed-point distribution functions. Indeed, Camp and Patey$^{17}$ recently studied criticality in three-dimensional model fluids with algebraically decaying attractive pair interactions varying as $-1/r^{3+\sigma}$ with exponent values $\sigma=3,1,$ and $0.1$, via grand canonical Monte Carlo simulations. When they applied the BW method, they found that the order-parameter distribution for $\sigma = 3$ could be well matched to the Ising fixed-point function; this is, in fact, consistent with the renormalization group theory conclusion that for $\sigma=3$ the system belongs to the Ising universality class.$^{26,27}$ 

On the other hand, they observed that for $\sigma=1$ and $0.1$ the order-parameter distributions could {\em not} be matched to the Ising fixed-point function so successfully. These observations might well have been anticipated since RG theory indicates that both these systems should behave classically,$^{26,27}$ i.e., display van der Waals critical exponents $(\alpha=0,\beta=\frac{1}{2},\gamma=1,\nu=\frac{1}{2})$.$^{23}$ Camp and Patey then tried to match the computed distribution functions to the approximately known classical fixed-point function$^{28}$ but with no success. They attributed the failure of the matching to lack of a precisely known classical fixed-point function; but it could well be due to the more singular corrections-to-scaling associated with slowly decaying power-law interactions or to the failure of a fortunate but seemingly accidental cancelation of such corrections found in Ising systems as explained below. Indeed, even if the fixed point function were accurately known, unresolved questions about the general validity and effectiveness of the Bruce-Wilding analysis would still remain, especially for fluids whose behavior deviates significantly from particle-hole symmetry such as polymer solutions and ``Coulombic'' vs.\ ``solvophobic'' electrolytes.$^{29}$ These questions will be addressed below.

One significant and potentially serious aspect of the BW approach has come to the fore only recently.$^{30,31}$ This pertains to the presence of a Yang-Yang anomaly$^{32}$ in the bulk critical behavior of the system of interest. To be explicit, suppose $\mu_{\sigma}(T)$ is the chemical potential on the phase boundary at and below $\Tc\,$: then a Yang-Yang anomaly is represented by the divergence (or, more generally, corresponding singular behavior) of the second derivative, $(d^{2}\mu_{\sigma}/dT^{2})$ when $T\rightarrow\Tc-$. The strength and sign of a Yang-Yang anomaly is conveniently measured$^{30,31}$ by ${\cal R}_{\mu}$, the limiting ratio of $\tilde{C}_{\mu}(T)\equiv -T(d^{2}\mu_{\sigma}/dT^{2})$ to the constant-volume specific heat, $C_{V}(T;\rho=\rhoc)$ (or, more generally, to its singular part).

It must be emphasized that although Yang-Yang anomalies vanish identically by construction in simple fluid models with an underlying gas-liquid (or `particle-hole') symmetry, such as the standard lattice-gases in which $\mu_{\sigma}(T)$ is always an analytic function through $\Tc$, they must appear in general, as Yang and Yang$^{32}$ argued nearly 40 years ago. Furthermore, both experimental data$^{33}$ and unequivocal simulation evidence --- for the hard-core square-well (HCSW) fluid$^{34,35}$ and the restricted primitive model (RPM) electrolyte$^{35}$ --- point to significant nonvanishing magnitudes of the Yang-Yang ratios ${\cal R}_{\mu}$ (even though the experimental situation is not yet fully transparent$^{36}$). In addition, an exactly soluble class of ``compressible cell gas'' models explicitly exhibits Yang-Yang anomalies.$^{30}$

But how do such anomalies impact the BW method? The answer is that BW specifically failed to allow for the possible mixing of the pressure $p$ into the scaling fields $\tilde{h}$ and $\tilde{t}$ via a term proportional to $p-p_{\mbox{\scriptsize c}}$. It transpires, however, that this previously neglected feature$^{21,22}$ of a ``full'' scaling theory is essential$^{30,31}$ to account for the presence of a Yang-Yang anomaly in the bulk critical behavior. Accordingly, one of the aims of the analysis presented below is to investigate how the introduction of pressure mixing, which entails two further, independent mixing coefficients, affects the precision and reliability of the BW method.

To tackle these issues we will invoke a rather thorough analysis of the full bulk scaling theory that incorporates pressure mixing and is applicable to nonsymmetric systems.$^{30}$ Then we will appeal to a recent systematic extension of the theory to systems of finite-size.$^{37}$ However, even to critique the BW theory in the absence of pressure mixing, it has proved necessary to formulate the BW matching procedures more precisely. It seems, indeed, that neither concrete details of actual implementations of the BW approach nor an explicit analysis of the relevant finite-size effects are available in the literature. Accordingly, in section III below, we revisit the BW approach (without pressure mixing) from first-principles and describe the essential details in a systematic and explicit manner.

On this basis, but subject to at least one serious proviso, we confirm the principal claims.$^{2,3}$ In particular, the effective {\em finite-size critical density}, $\rhoc(L)$, defined by BW approaches $\rhoc$ like $1/L^{(1-\alpha)/\nu}$ when $L\rightarrow\infty$ (where $\alpha$ and $\nu$ are critical exponents for the specific heat and the correlation length$^{22}$). Likewise, the corresponding {\em effective critical temperature}, $\Tc(L)$, approaches $\Tc$ as $1/L^{(1+\theta)/\nu}$ (where $\theta$ is the leading even correction-to-scaling exponent$^{23}$). Armed with this information, effective extrapolations to $L=\infty$ may be undertaken. In addition to verifying these convergence laws, we obtain the corresponding leading amplitudes of the deviations, $\Delta\Tc(L)$ and $\Delta\rhoc(L)$, in terms of parameters that are known or given.

We then ask: Does pressure mixing change these results and, if so, how? Can the approach be modified readily to allow for the changes? In answering these questions, we use the finite-size scaling form for the {\em canonical} free energy, $f(\rho,T)$, derived in ref 37. We conclude, in fact, that pressure mixing will generate a {\em new term} in $\rhoc(L)$ varying as $1/L^{2\beta/\nu}$ (where $\beta$ is the coexistence-curve exponent$^{23}$) that dominates the previously derived $1/L^{(1-\alpha)/\nu}$ term for large $L$. For Ising type $(d$$\,=\,$$3)$-dimensional fluids one has $2\beta/\nu\simeq 1.03$ while $(1-\alpha)/\nu\simeq1.41$ so the difference exponent is only about 0.38.$^{23}$

The amplitude of this $L^{-2\beta/\nu}$ term is proportional to the coefficient $j_{2}$ that specifies the degree to which the pressure mixes into the ordering field $\tilde{h}$: see eq 3 below. Thus, when pressure mixing is negligible, i.e., $j_{2}$ is sufficiently small, the BW estimation procedure for $\rhoc$ (that assumes the asymptotic $1/L^{(1-\alpha)/\nu}$ behavior) remains reasonable. However, one should expect the dominant $1/L^{2\beta/\nu}$ term to play a more significant role in highly asymmetric fluids like the RPM where pressure mixing seems likely to be larger. Furthermore, these two terms may compete which could well yield potentially misleading nonmonotonic behavior. In the case of $\Tc(L)$, our analysis indicates that pressure mixing does not alter the leading $1/L^{(1+\theta)/\nu}$ term although its amplitude changes if $j_{2}\neq 0$; however, that does not invalidate the BW approach.

At this point we should point out that to fulfill the need for {\em bias-free} methods of assessing fluid criticality and, in particular, to convincingly determine universality class without prior assumptions, novel approaches have recently been devised, analyzed, and implemented.$^{34,35,37,38}$ These techniques rest on the calculation of certain special finite-size loci in the $(\rho,T)$ plane, notably the $k$-loci$^{34}$ and, especially, the $Q$-loci,$^{38}$ $\rho_{Q}(T;L)$, and the examination of various density fluctuation moments as functions of $T$ and $L$ on these loci. In this way the Ising character of criticality in the RPM electrolyte has been established,$^{35,38}$ and precise and, apparently, rather reliable estimates of $\Tc$ and $\rhoc$ have been found for the model.

However, these new methods require more and more careful computation! Accordingly, for ``every-day'' or exploratory applications where Ising behavior may be reasonably presupposed, it is reasonable to ask what kind of error one should expect in applying the convenient and relatively economical BW recipe to estimate the critical points of new model fluids. Indeed, the BW method does seem to give reliable estimates of $\Tc$ when the universality class of the model under investigation is of Ising character. Some concrete evidence for this conclusion is presented in Table 1, which gives estimates for the HCSW fluid and the RPM. One sees that for the rather symmetric HCSW fluid (with a very small Yang-Yang anomaly,$^{35}$ ${\cal R}_{\mu}\simeq -0.04$) the central BW estimate for $\Tc$ is only same 1 part in $10^{4}$ higher than the unbiased value; indeed, the quoted uncertainties overlap. For the RPM (with a highly asymmetric coexistence curve and a significant Yang-Yang anomaly,$^{35}$ ${\cal R}_{\mu}\simeq +0.2_{6}$) the uncertainties no longer overlap; but the BW estimate is lower by less than 1 part in $10^{3}$. On the other hand, the critical densities provided by the BW method and their apparent precision are in much poorer agreement with the new unbiased estimates; for both models the BW values are higher, by $1\%$ and $6\%$, respectively. One may suspect that a similar situation will prevail more generally; but further precise simulations, for example, for polymer solution models, are needed to confirm such a verdict.

The balance of this article is organized as follows: In section II we briefly recapitulate, for ease of reference and to define notation, the basic scaling theory with pressure mixing.$^{30}$ The vital details of the BW method and concrete procedures for matching the distribution functions are formulated in section III. The limiting behavior of the basic estimators $\rhoc(L)$ and $\Tc(L)$ is then discussed pointing out the fortunate accident (noted but not stressed by the original authors) that seems essential to its practical success. Section IV addresses the modifications of the BW analysis that allow for pressure mixing. In section V a brief summary and discussion is provided.

\ssection{\hspace{-.27in}.\hspace{0.1in}Full Scaling Theory for Fluids}
\subsection{\hspace{-.24in}\hspace{0.1in}Bulk thermodynamics}
\indent

Following refs 30 and 31 we consider a single-component fluid with a critical point at $p_{\mbox{\scriptsize c}}$, $\mu_{\mbox{\scriptsize c}}$, and $\Tc$ and with a critical density $\rhoc$. Convenient dimensionless variables near the critical point are then 
 \begin{equation}
  \check{p} \equiv (p-p_{\mbox{\scriptsize c}})/\rhoc k_{\mbox{\scriptsize B}}\Tc, \hspace{.2in} \check{\mu} \equiv (\mu-\mu_{\mbox{\scriptsize c}})/k_{\mbox{\scriptsize B}}\Tc, \hspace{0.2in} t \equiv (T-\Tc)/\Tc, \label{ch5-eq5.5}
 \end{equation}
where $k_{\mbox{\scriptsize B}}$ is Boltzmann's constant. Neglecting terms beyond linear order$^{31}$ the {\em nonlinear scaling fields}, $\tilde{p}$, $\tilde{h}$, and $\tilde{t}$ may be written
 \begin{eqnarray}
  \tilde{p} & = & \check{p}-k_{0}t-l_{0}\check{\mu} + \cdots,  \label{ch5-eq5.2} \\
  \tilde{h} & = & \check{\mu}-k_{1}t-j_{2}\check{p} + \cdots, \label{ch5-eq5.3} \\
  \tilde{t} & = & t-l_{1}\check{\mu}-j_{1}\check{p} + \cdots,   \label{ch5-eq5.4}
 \end{eqnarray}
where, for future reference we mention that in the bulk or thermodynamic limit the choices in eq 1 leads to $l_{0}=1$ while $k_{0}$ is related to the critical point entropy. [See eq 3.22 of ref 30.] In these terms a general scaling hypothesis asserts that the thermodynamics near criticality can be described, at least asymptotically, by$^{31}$
  \begin{equation}
   \tilde{p} \approx Q|\tilde{t}|^{2-\alpha}W_{\pm}(U\tilde{h}/|\tilde{t}|^{\Delta},U_{4}|\tilde{t}|^{\theta},U_{5}|\tilde{t}|^{\theta_{5}}),
  \end{equation}
with $\pm$ corresponding to $\tilde{t}\gtrless 0$, where $\Delta =\beta+\gamma$ is the gap exponent while $\alpha$, $\beta$ and $\gamma$ are the standard exponents.$^{23}$ As mentioned above, $\theta$ is the leading even correction-to-scaling exponent while $\theta_{5}$ is the corresponding leading odd exponent. The coefficients $Q$, $U$, $U_{4}$ and $U_{5}$ are nonuniversal amplitudes while, when suitably normalized,$^{31}$ $W_{\pm}$ are two branches of a universal scaling function. The nonuniversal irrelevant scaling amplitudes, $U_{4}$ and $U_{5}$, will, in general, depend smoothly on the variables $t$, $\check{p}$, $\check{\mu}$ or $\tilde{p}$, $\tilde{h}$ and $\tilde{t}$.$^{23,31}$ Note that the BW assumption corresponds to setting $j_{1}=j_{2}=0$ (which does not, in and of itself, relate to any special symmetry); thus their analysis requires the determination of only the two mixing coefficients $k_{1}$ and $l_{1}$.

In line with the original BW analysis, it is appropriate to define two relevant critical densities or ``operators,'' ${\cal M}$ and ${\cal E}$, conjugate to the ordering field $\tilde{h}$ and the thermal scaling field $\tilde{t}$ in the underlying Hamiltonian. In the context of an Ising ferromagnet these may be identified as the fluctuating magnetization and the fluctuating energy density (or, more precisely, as the deviations from the respective mean values at criticality). Thus we may obtain the mean values of these basic scaling densities via
   \begin{equation}
  \langle\delta {\cal M}\rangle = \langle{\cal M}-{\cal M}_{\mbox{\scriptsize c}}\rangle= \left(\frac{\partial \tilde{p}}{\partial \tilde{h}}\right)_{\tilde{t}},\hspace{.3in} \langle\delta {\cal E}\rangle = \langle{\cal E}-{\cal E}_{\mbox{\scriptsize c}}\rangle= -\left(\frac{\partial \tilde{p}}{\partial \tilde{t}}\right)_{\tilde{h}},   \label{ch5-eq5.6}
 \end{equation}
where the subscripts denote mean values evaluated at criticality. (Note that in the previous analysis$\,^{31}$ the notation adopted was $\tilde{\rho}=\langle\delta {\cal M}\rangle$ and $\tilde{s}=-\langle\delta {\cal E}\rangle$. We may also define the thermodynamic reduced number density $\check{\rho}$ and energy density $\check{u}$ via$^{31}$
 \begin{equation}
  \check{\rho} \equiv \left(\frac{\partial \check{p}}{\partial \check{\mu}}\right)_{t}, \hspace{0.3in} \check{u} \equiv -\left(\frac{\partial \check{p}}{\partial t}\right)_{\mu}.  \label{ch5-eq5.7}
 \end{equation}
These are related to the number density, $\rho = N/V$, and the total entropy, $S$, by
 \begin{equation}
  \check{\rho} \equiv \check{\rhoc} + \Delta\check{\rho} = \frac{\rho}{\rhoc}, \hspace{0.2in}  \check{u} \equiv \check{u}_{\mbox{\scriptsize c}} + \Delta\check{u} = -\frac{S}{V\rhoc k_{\mbox{\scriptsize B}}},  \label{ch5-eq5.7a}
 \end{equation}
where, as previously, $N$ is the number of particles in the system and $V$ the volume.

Now the relevant critical densities, $\langle\delta{\cal M}\rangle$ and $\langle\delta{\cal E}\rangle$, can be expressed as {\em nonlinear combinations} of the reduced energy and number densities: after some algebra one finds$^{31}$
 \begin{eqnarray}
  \langle\delta {\cal M}\rangle & \approx & \frac{(1-j_{1}k_{0})\Delta\check{\rho} - (l_{1}+j_{1}l_{0})\Delta \check{u}}{K - (j_{2}+j_{1}k_{1})\Delta\check{\rho} + (j_{1}+j_{2}l_{1})\Delta \check{u}}, \label{ch5-eq5.8}  \\
  \langle\delta {\cal E}\rangle & \approx & \frac{(1-j_{2}l_{0})\Delta \check{u} - (k_{1}+j_{2}k_{0})\Delta\check{\rho}}{K-(j_{2}+j_{1}k_{1})\Delta\check{\rho} + (j_{1}+j_{2}l_{1})\Delta \check{u}},  \label{ch5-eq5.9}
 \end{eqnarray}
where the constant
 \begin{equation}
  K = 1-k_{1}l_{1}-j_{1}k_{0}-j_{2}l_{0}-j_{2}k_{0}l_{1}-j_{1}k_{1}l_{0}, \label{ch5-eq5.10}
 \end{equation}
reduces to $1-k_{1}l_{1}$ in the absence of pressure mixing. Note that even though the relevant scaling fields are considered only up to linear order, the presence of the {\em linear} pressure-mixing makes the mean order-parameter $\langle{\cal M}\rangle$ and critical energy density $\langle{\cal E}\rangle$ {\em nonlinear} combinations of the number density $\check{\rho}$ and the energy density $\check{u}$. This is in contrast to the linear forms assumed by BW for ${\cal M}$ and ${\cal E}$ in terms of the fluctuating number and energy densities:$^{2,3}$ however, in the absence of pressure mixing ( i.e., when $j_{1}=j_{2}=0$) the two mean densities reduce to
 \begin{equation}
  \langle\delta {\cal M}\rangle = \frac{\Delta\check{\rho}-l_{1}\Delta\check{u}}{1-k_{1}l_{1}}, \hspace{0.3in} \langle\delta {\cal E}\rangle = \frac{\Delta\check{u} - k_{1}\Delta\check{\rho}}{1-k_{1}l_{1}},  \label{ch5-eq5.11}
 \end{equation}
in accord with the BW formulation.

Even in the presence of pressure-mixing, one can expand eqs 9 and 10 near the critical point up to linear order to obtain
 \begin{eqnarray}
  \langle\delta {\cal M}\rangle & \approx & K^{-1}[(1-j_{1}k_{0})\Delta\check{\rho}-(l_{1}+j_{1}l_{0})\Delta \check{u}],  \label{ch5-eq5.12} \\
  \langle\delta {\cal E}\rangle & \approx & K^{-1}[(1-j_{2}l_{0})\Delta \check{u} - (k_{1}+j_{2}k_{0})\Delta\check{\rho}].  \label{ch5-eq5.13}
 \end{eqnarray}
These forms are consistent with the expressions given by BW except for the modification of the coefficients by the pressure-mixing coefficients $j_{1}$ and $j_{2}$.

\subsection{\hspace{-.24in}\hspace{0.1in}Finite-size scaling}
\indent

To discuss the analysis of simulation data we need an appropriate, ``complete'' finite-size scaling formulation. To that end we follow ref 37 and first introduce the dimensionless size-scaled temperature and ordering field variables
  \begin{equation}
   w_{L} = a_{\cal E}\tilde{t}L^{1/\nu}, \hspace{0.3in} z_{L} = a_{\cal M}\tilde{h}L^{\Delta/\nu},
  \end{equation}
and the leading even and odd correction variables
 \begin{equation}
   z_{L4} = a_{4}L^{-\theta/\nu} \hspace{0.3in} \mbox{and}\hspace{0.3in} z_{L5} = a_{5}L^{-\theta_{5}/\nu}.
 \end{equation}
Here $\nu$ is the correlation length exponent while $a_{\cal E}$, $a_{\cal M}$, $a_{4}$, and $a_{5}$ are nonuniversal metrical factors (depending only weakly on $p$, $\mu$, $T$ and $L$) of dimensions conjugate to the powers of $L$ appearing in the definitions of $w_{L}$, etc. The basic scaling hypothesis eq 5 is then extended to
  \begin{equation}
   \rhoc\tilde{p} \approx L^{-d} Y(w_{L},z_{L};z_{L4},z_{L5}).
  \end{equation}
See eq 2.2 of ref 37.$^{39}$ Apart from possible $L$-dependences of the scaling field$^{37}$ $\tilde{p}$, $\tilde{h}$ and $\tilde{t}$, which, however, have no effects to the orders that will be relevant here,$^{37}$ the definitions in eqs 2, 3, and 4 still hold. The scaling function $Y(w,z;z_{4},z_{5})$ should be universal (when suitably normalized) but {\em must} depend on the boundary conditions and, indeed, on the shape of the system domain. Accordingly, for concreteness we will presuppose cubic simulation boxes of edge length $L$ with periodic boundary conditions. Since in a finite system the grand canonical pressure must be related analytically to the other fields, the scaling function $Y(x_{L},\cdots)$ should be analytic for small values of all its arguments.$^{37}$

The bulk limit may now be obtained formally by setting $L=1/|a_{\cal E}\tilde{t}|^{\nu}$ and letting $L\rightarrow\infty$. The form of eq 5 is then recaptured provided the {\em hyperscaling relation} $d\nu = 2-\alpha$ is accepted.$^{40}$ One further finds $Q=|a_{\cal E}|^{2-\alpha}/\rhoc$, which is dimensionless, $U=a_{\cal M}/|a_{\cal E}|^{\Delta}$, $U_{4}=a_{4}|a_{\cal E}|^{\theta}$, and $U_{5}=a_{5}|a_{\cal E}|^{\theta_{5}}$, while $W_{\pm}(z;z_{4},z_{5})=Y(\pm 1,z;z_{4},z_{5})$.

Now BW focus on the joint distribution function, $P_{L}({\cal M},{\cal E})$, of the critical densities ${\cal M}$ and ${\cal E}$ and suppose this has an appropriate universal scaling form. Without doubt the mean values, $\langle \delta {\cal M}\rangle_{L}$ and $\langle {\cal E}\rangle_{L}$, in a finite system will still be given by the previous thermodynamic first derivatives --- see eq 6 --- provided $\tilde{p}$ is replaced by the finite-size expression eq 17 for $\tilde{p}(p,\mu,T;L)$.$^{37}$ However, the mean values clearly do {\em not} define the distribution function $P_{L}({\cal M},{\cal E})$. On the other hand, from the successive derivatives $(\partial^{m+n}\tilde{p}/\partial\tilde{h}^{m}\partial\tilde{t}^{n})$ one can extract, in a standard way, the corresponding moments $\langle (\delta {\cal M})^{m}(\delta {\cal E})^{n}\rangle_{L}$ for all $(m,n)$ as functions of $\tilde{h}$ and $\tilde{t}$ (or, equivalently, as functions of $p$, $\mu$, and $T$). Then, from the full set of moments one may, at least in principle, reconstruct the corresponding distribution function that is desired.$^{41}$ Furthermore, it is clear that all the scaling properties of the set of moments will be inherited by the distribution function.

If we introduce the scaled number and energy fluctuation variables via
 \begin{equation}
   x_{L} = \delta {\cal M}L^{\beta/\nu}/a_{\cal M} \hspace{0.3in} \mbox{and} \hspace{0.3in} y_{L}=\delta {\cal E}L^{(1-\alpha)/\nu}/a_{\cal E},
 \end{equation}
it follows that we may conclude
 \begin{equation}
  P_{L}({\cal M},{\cal E}) \approx {\cal N}_{L}{\cal P}(x_{L},y_{L};w_{L},z_{L};z_{L4},z_{L5}),
 \end{equation}
where the normalization constant ${\cal N}_{L}$ can play no special role. Apart from notation and the neglect of the leading odd correction variable $z_{L5}$, this is precisely the finite-size scaling ansatz advanced by Wilding.$^{7}$ (Of course we ourselves have, here and above, neglected all higher order correction variables $z_{Lk}$ for $k>5$.$^{23,31,37}$) It is also clear that the scaling function ${\cal P}(x_{L},\cdots)$ should be universal. At bulk criticality, where $\tilde{h}=\tilde{t}=0$, so that $w_{L}=z_{L}=0$, we can then write
 \begin{equation}
  P_{L}^{\mbox{\scriptsize c}}({\cal M},{\cal E}) \approx {\cal N}_{L} {\cal P}_{\mbox{\scriptsize c}}( \delta {\cal M}L^{\beta/\nu}/a_{\cal M},\delta {\cal E}L^{(1-\alpha)/\nu}/a_{\cal E}; a_{4}/L^{\theta/\nu},a_{5}/L^{\theta_{5}/\nu})
 \end{equation}
where ${\cal P}_{\mbox{\scriptsize c}}(x_{L},\cdots)$ embodies the universal, statistically scale-invariant critical density fluctuations characteristic of a specific universality class but also incorporates the leading inverse powers of $L$ that must play a significant role in systems that are not very large.

\ssection{\hspace{-.27in}.\hspace{0.1in}Analysis of the Bruce-Wilding Method}
\indent

The aim of the BW method is to match the observed fluctuation data for the particle number density, $\delta\check{\rho}$, and energy density, $\delta\check{u}$, to the expected near-critical ordering density distribution, that follows from
 \begin{equation}
  p_{L,{\cal M}}({\cal M}) = \int P_{L}({\cal M},{\cal E}) d{\cal E},
 \end{equation}
on the basis of the scaling ansatz eq 20, and likewise to the corresponding critical energy density distribution, $p_{L,{\cal E}}({\cal E})$. In essence the matching is to be accomplished (i) by suitably choosing the two mixing parameters $l_{1}$ and $k_{1}$ in the expressions
 \begin{equation}
  \delta{\cal M} \propto \delta\check{\rho} - l_{1}\delta\check{u}, \hspace{0.3in} \delta{\cal E} \propto \delta\check{u}-k_{1}\delta\check{\rho},  
 \end{equation}
(that underlie eq 12 for the mean values) and (ii) by determining $L$-dependent approximations, $\Tc(L)$, $\rhoc(L)$ and $u_{\mbox{\scriptsize c}}(L)$, for the bulk values $\Tc$, $\rhoc$ and $u_{\mbox{\scriptsize c}}$. Appropriate extrapolations on $L$ should then yield optimal estimates for $\Tc$ and $\rhoc$.

The universal limiting $(L\rightarrow\infty)$ critical distributions, which we will call $p_{\cal M}^{\ast}(x)$ and $p_{\cal E}^{\ast}(y)$, where $x$ and $y$ are defined as in eq 18 (but with the subscripts $L$ dropped for brevity) are assumed known {\em a priori}. Indeed, fluid-magnet universality implies that a critical fluid order parameter distribution, $p_{L,{\cal M}}^{\mbox{\scriptsize c}}({\cal M})$, should, when $L\rightarrow\infty$, precisely match the universal function $p_{\cal M}^{\ast}(x)$ appropriate to the Ising universality class; but this can be found independently by careful Monte Carlo simulations of Ising models.$^{8,24,25}$ Figure 1 shows this distribution as obtained by Wilding and M\"{u}ller$^{6}$ via simulations of the $(d$$\,=\,$$3)$-dimensional Ising model. The distribution function has been normalized to unit integrated weight while the nonuniversal amplitude $a_{\cal M}$ is chosen so that the distribution has unit variance. Evidently $p_{\cal M}^{\ast}(x)$ has two symmetrical peaks at, say, $x=\pm x^{\ast}$ of equal height $p_{\mbox{\scriptsize max}}^{\ast\cal M}$, where this value and $x^{\ast}$ should be universal: see Fig.\ 1. The same approach applies to the energy distribution $p_{L,{\cal E}}^{\mbox{\scriptsize c}}({\cal E})$: this should match $p_{\cal E}^{\ast}(y)$ which may also be found numerically:$^{8}$ see Fig.\ 2. This distribution, which is unimodal but {\em asymmetric}, has been normalized similarly by choice of $a_{\cal E}$. Again the location of the peak at, say, $y=y^{\ast}$, and its height, $p_{\mbox{\scriptsize max}}^{\ast\cal E}$, must be universal: see Fig.\ 2. Other universality classes, such as, e.g., the XY or Heisenberg classes, etc. may be expected to have broadly similar distributions but with distinct values of $x^{\ast}$, $y^{\ast}$, etc.

Now let us first consider the BW theory, in the absence of pressure mixing, and formulate their matching procedures in an explicit manner.

\subsection{\hspace{-.24in}\hspace{0.1in}Matching conditions}
\indent

The finite-size ordering distribution, $p_{L,{\cal M}}({\cal M})$, can be obtained computationally by transforming the joint distribution $p_{L}(\check{\rho},\check{u})$ as observed in a simulation (where here we regard $\check{\rho}$ and $\check{u}$ as the {\em fluctuating} finite-size values of density and energy), by using eqs 21 and 22, while simultaneously transforming the chemical potential $\mu$ and the temperature $T$ by introducing the (as yet unknown) mixing coefficient $l_{1}$. As indicated, when $L\rightarrow\infty$, the resulting function, $p_{L,{\cal M}}({\cal M})$, should match the universal distribution $p^{*}_{{\cal M}}(x)$ in terms of the variable
 \begin{equation}
   x=L^{\beta/\nu}({\cal M}-{\cal M}_{\mbox{\scriptsize c}})/a_{\cal M}.  \label{ch5-eq5.19}
 \end{equation}
Using the (known) critical exponents $\beta$ and $\nu$ for the universality class in question, matching $p_{L,{\cal M}}({\cal M})$ to the universal function $p^{*}_{{\cal M}}(x)$ at some fixed $L$ is to be implemented by choosing five fitting parameters, which, following BW, we take as the estimators $\Tc(L)$, $\mu_{\mbox{\scriptsize c}}(L)$, $l_{1}(L)$, $a_{{\cal M}}(L)$ and ${\cal M}_{\mbox{\scriptsize c}}(L)$ since, when $L\rightarrow \infty$, these should approach the corresponding bulk values.

Wilding has given no details of how he actually performs the matching that he implements:$^{9}$ this leaves some ambiguity when it comes to estimating finite-size corrections and studying the effects of pressure mixing. Accordingly, for precision we consider the following formal procedure: {\bf (a)} At finite $L$ and fixed $T\lesssim \Tc$ [and for values of $\mu$ not too far from $\mu_{\sigma}(T)$] the measured distribution $p_{L,{\cal M}}({\cal M})$ will normally be two-peaked. By adjusting $\mu$ to, say, $\mu^{(1)}(L)$ the peaks can be made of equal heights; But, in general they will not be symmetrically disposed with respect to the intervening minimum. It should be possible to achieve this by {\bf (b)} adjusting $l_{1}$ and, generally, re-adjusting $\mu$ to obtain values $\mu^{(2)}(L)$ and $l_{1}^{(2)}(L)$; {\bf (b1)} the position of the minimum then identifies an estimate ${\cal M}_{\mbox{\scriptsize c}}^{(2)}(L)$ and {\bf (b2)} one can then satisfy the universal peak-placement condition
 \begin{equation}
  x_{L}^{\pm} \equiv a_{{\cal M}}^{-1}(L)\: L^{\beta/\nu}({\cal M}_{\pm}^{(2)}-{\cal M}_{\mbox{\scriptsize c}}) = \pm x^{*},  \label{ch5-eq5.20}
 \end{equation}
for the peaks at ${\cal M}_{\pm}^{(2)}(L)$ by adjusting $a_{{\cal M}}$ to, say, $a_{{\cal M}}^{(2)}(L)$: At this point, in general, the peak heights will not satisfy the universal relation desired, namely,
 \begin{equation}
  p_{L}^{\pm} \equiv p_{L,{\cal M}}({\cal M}_{\pm}^{(2)}) = p_{{\cal M}}^{*}(\pm x^{*}) \equiv p^{\ast\cal M}_{\mbox{\scriptsize max}}.  \label{ch5-eq5.21}
 \end{equation}
Nor will the height of the minimum satisfy the corresponding relation
 \begin{equation}
  p_{L}^{0} \equiv p_{L,{\cal M}}({\cal M}_{\mbox{\scriptsize c}}^{(2)}) = p_{{\cal M}}^{*}(0) \equiv p^{\ast\cal M}_{\mbox{\scriptsize min}}.  \label{ch5-eq5.22}
 \end{equation}
See Fig.\ 1. {\bf (c)} To satisfy these two conditions it should (ignoring finite-size and pressure mixing corrections) suffice to vary the temperature $T$. Changing $T$ should yield a series of further estimates $\mu^{(3)}(L)$, $\mu^{(4)}(L)$, $\cdots$, for $\mu_{\mbox{\scriptsize c}}(L)$ and, likewise, for the other parameters. Unless precision is high, both eqs 25 and 26 should yield the same set of five fitting parameters $\Tc(L)$, $\mu_{\mbox{\scriptsize c}}(L)$, $l_{1}(L)$, $a_{{\cal M}}(L)$, and ${\cal M}_{\mbox{\scriptsize c}}(L)$: significant differences, if they arise, could be a consequence of either finite-size or pressure-mixing effects (or the `known' universal distribution could be in error). However, practical experience for Ising-type systems suggests that satisfactory fits within the numerical uncertainties can be obtained fairly readily; but, even this might change if data of significantly higher precision became available.

To determine the remaining fitting parameters for the energy, namely, ${\cal E}_{\mbox{\scriptsize c}}(L)$, $a_{{\cal E}}(L)$, and $k_{1}(L)$, one then considers the variable 
 \begin{equation}
  y = L^{(1-\alpha)/\nu}({\cal E}-{\cal E}_{\mbox{\scriptsize c}})/a_{\cal E},  \label{ch5-eq5.23}
 \end{equation}
in order to match the data for the finite-size energy distribution, $p_{L,{\cal E}}({\cal E})$, to the known fixed-point function $p_{{\cal E}}^{*}(y)$ of character as shown in Fig.\ 2.

We also remark that if pressure mixing is absent, i.e., $j_{1}=j_{2}=0$, the fitting values $\rhoc(L)$ and $u_{\mbox{\scriptsize c}}(L)$ may also be determined by using eq 12. Notice then that the mixing parameters, $l_{0}(L)$, and $k_{0}(L)$ are simply equal to $\check{\rho}_{\mbox{\scriptsize c}}(L)$ and $-\check{u}_{\mbox{\scriptsize c}}(L)$, respectively, as follows from the scaling analysis.$^{31,37}$

\subsection{\hspace{-.24in}\hspace{0.1in}{\boldmath $L$}-dependence of estimators for {\boldmath $\Tc$} and {\boldmath $\mu_{\mbox{\scriptsize c}}$}}
\indent

To elucidate the $L$-dependence of the estimators $\Tc(L)$, $\rhoc(L)$, etc., determined from the fits at fixed $L$ as set out above, let us integrate the joint distribution eq 19 over the energy fluctuations to obtain the order distribution, $p_{L,{\cal M}}({\cal M})$. We may expand this distribution about the fixed point function $p_{\cal M}^{*}(x)$ as
 \begin{eqnarray}
  p_{L,{\cal M}}({\cal M}) & \approx & {\cal N}_{\cal M}[p_{{\cal M}}^{*}(x) + a_{{\cal E}}\: L^{1/\nu}\tilde{t}p_{t}^{*}(x)+ a_{4}\: L^{-\theta/\nu} p_{4}^{*}(x) \nonumber \\
  &  & \hspace{0.65in} +\: a_{\cal M}L^{\Delta/\nu}\tilde{h}p_{h}^{*}(x) + a_{5}L^{-\theta_{5}/\nu}p_{5}^{*}(x) + \cdots ],  \label{ch5-eq5.24}
 \end{eqnarray}
where $x$ is defined in eq 23 and the normalizating factor will play no role. The derivative scaling functions $p_{t}^{*}(x)$, $p_{4}^{*}(x)$, etc., should be universal. Note also that by the symmetry of the Ising (or similar) fixed point, the functions $p_{\cal M}^{*}(x)$, $p_{t}^{*}(x)$ and $p_{4}^{*}(x)$ are symmetric under $x \Leftrightarrow -x$, while $p_{h}^{*}(x)$ and $p_{5}^{*}(x)$, which describe the contributions from the ordering field and the odd corrections-to-scaling, respectively, must be antisymmetric.

The original BW approach assumed tacitly that throughout the matching procedure $\tilde{h}=0$ is maintained; it also ignored the odd correction-to-scaling contribution. In this case, the order distribution $p_{L,{\cal M}}({\cal M})$ becomes symmetric about ${\cal M}={\cal M}_{\mbox{\scriptsize c}}$. This will, in fact, be true for systems displaying an Ising-type symmetry. However, for asymmetric (e.g., fluid) systems, the ordering field $\tilde{h}$ at bulk $\Tc$ and $\rhoc$ will depend on the system size $L$. In fact, one can show$^{37}$ that in this case the ordering field $\tilde{h}$ decays as
 \begin{equation}
  \tilde{h} = \tilde{a}_{h}/L^{(1-\alpha+\gamma)/\nu} + \cdots, \label{ch5-eq5.24a}
 \end{equation}
where, in the absence of pressure mixing, the amplitude $\tilde{a}_{h}$ is proportional to the mixing coefficient $l_{1}$. Note also that $l_{1}$ generates a singular $|t|^{1-\alpha}$ term in the coexistence curve diameter.$^{21,22}$ M\"{u}ller and Wilding$\,^{8}$ also noticed this point in their study of an asymmetric binary polymer mixture and observed that the chemical potential deviation on the phase boundary decays with the same exponent $(1-\alpha+\gamma)/\nu$. The consequent presence of an {\em antisymmetric} contribution to $p_{L,{\cal M}}({\cal M})$, which varies as $L^{\Delta/\nu}\tilde{h} \propto L^{-(1-\alpha-\beta)/\nu}$ then makes it more difficult to match the order distribution to the {\em symmetric} fixed point function for $L$ not so large. This effect might cause some difficulties in applying the BW method to highly asymmetric fluids even if pressure mixing may be neglected.

Nevertheless, in order to elucidate the original BW method, let us initially ignore the contributions in eq 28 proportional to $p_{h}^{\ast}(x)$ and $p_{5}^{\ast}(x)$ that arise from the nonzero value of $\tilde{h}$ and the odd corrections-to-scaling. In that case, the order distribution $p_{L,{\cal M}}({\cal M})$ should be symmetric in ${\cal M}$ about the critical value ${\cal M}_{\mbox{\scriptsize c}}$. But then the rather precise collapse of $p_{L,{\cal M}}({\cal M})$ onto the fixed point function $p_{{\cal M}}^{*}(x)$ that has been achieved in practical simulations seems to require the {\em effective cancelation} of the contributions from the finite non-zero value of the scaling field $\tilde{t}$ by the leading even correction-to-scaling term: see the second and third terms in eq 28. Indeed, as noted by Wilding, Nicolaides and Bruce$^{7,24}$ the functional forms of the universal functions $p_{t}^{*}(x)$ and $p_{4}^{*}(x)$ do seem to be similar in the Ising universality class, so that such a cancelation is feasible! However, this seems to be no more than a fortunate accident with only limited numerical support; no firm evidence as to why this should be true or should hold for other universality classes has been offered. Nevertheless, if one accepts this observation as a reasonable approximation, the cancelation of the two contributions yields
 \begin{equation}
  a_{{\cal E}}\: L^{1/\nu}\:\tilde{t} + a_{4}\: L^{-\theta/\nu}\: {\cal R}^{\ast} \simeq 0, \label{ch5-eq5.25}
 \end{equation}
where ${\cal R}^{\ast}$ is the (approximately) $x$-independent ratio of $p_{4}^{*}(x)/p_{t}^{*}(x)$. The condition $\tilde{h} = 0$, which was accepted in the original BW method, leads via the scaling field definition eq 3, to $\check{\mu} = k_{1}t$ where we are, here, neglecting $j_{2}$. Substituting and using eq 4 (with $j_{1}=0$) leads to
 \begin{equation}
   t_{\mbox{\scriptsize c}}(L) \equiv \frac{\Tc(L)-\Tc^{\infty}}{\Tc^{\infty}} \approx - \frac{a_{4}{\cal R}^{\ast}L^{-(\theta+1)/\nu}}{(1-k_{1}l_{1})a_{{\cal E}}},  \label{ch5-eq5.27}
 \end{equation}
where $\Tc^{\infty}$ is the true critical temperature. The exponent $(1+\theta)/\nu$ here confirms the arguments of Wilding, Nicolaides and Bruce.$^{7,24}$ For a nonzero value of $k_{1}$, it also follows that the chemical potential estimator scales in the same way as $\Tc(L)$. We thus obtain explicitly
 \begin{equation}
  \mu_{\mbox{\scriptsize c}}(L)-\mu_{\mbox{\scriptsize c}}^{\infty} \approx -\frac{k_{1}a_{4}{\cal R}^{\ast}k_{\mbox{\scriptsize B}}\Tc}{(1-k_{1}l_{1})a_{\cal E}}L^{-(\theta +1)/\nu},  \label{ch5-eq5.28}
 \end{equation}
where $\mu_{\mbox{\scriptsize c}}^{\infty}$ is the bulk critical chemical potential. Finally, recognizing eq 29 for $\tilde{h}(L)$, turns out not to change the leading behavior in the temperature and chemical potential estimators, since, at least for $(d$$\,=\,$$3)$-dimensional Ising-type systems, we have $(1-\alpha+\gamma)\simeq 2.13 > (1+\theta)\simeq 1.5$. But it must be emphasized again that this conclusion relies upon the surely approximate proportionality of $p_{t}^{*}(x)$ to $p_{4}^{*}(x)$.

\subsection{\hspace{-.24in}\hspace{0.1in}Convergence of the energy and density estimators}
\indent

To study estimators for the true critical density, $\rhoc$ ($\equiv \rhoc^{\infty}$), and energy density, $u_{\mbox{\scriptsize c}}$ ($\equiv u_{\mbox{\scriptsize c}}^{\infty}$), we consider the following equations derived by inverting eq 12, namely,
 \begin{equation}
  \Delta\check{\rho} = \langle\delta {\cal M}\rangle + l_{1}\langle \delta {\cal E}\rangle, \hspace{.3in} \Delta\check{u} = \langle\delta {\cal E}\rangle + k_{1} \langle\delta {\cal M}\rangle.  \label{ch5-eq5.29}
 \end{equation}
In order to obtain $\check{\rho}_{\mbox{\scriptsize c}}(L)$ and $\check{u}_{\mbox{\scriptsize c}}(L)$ we must thus first determine the behavior of ${\cal M}_{\mbox{\scriptsize c}}(L)$ and ${\cal E}_{\mbox{\scriptsize c}}(L)$. Wilding and M\"{u}ller$^{6,7}$ argued that the critical value of the ordering operator, namely, ${\cal M}_{\mbox{\scriptsize c}}(L)$, should not depend on the system size $L$, since the asymptotic order distribution $p_{L,{\cal M}}({\cal M})$ will be symmetric in ${\cal M}-\langle {\cal M}\rangle$ which fixes ${\cal M}_{\mbox{\scriptsize c}}$. However, this cannot be strictly true if one recognizes the contributions from the nonzero value of $\tilde{h}$ and the leading odd correction-to-scaling term. As mentioned, these contributions, proportional to $p_{h}^{*}(x)$ and $p_{5}^{*}(x)$, respectively, are antisymmetric in $x$. Therefore, for any finite $L$, a perfect collapse of $p_{L,{\cal M}}({\cal M})$ onto the symmetric, fixed point distribution $p_{\cal M}^{*}(x)$ is not, in general, possible. Nevertheless, let us suppose that an optimal collapse of data has been achieved (say, by some least-squares procedure) yielding a best fit for the various estimators. How does ${\cal M}_{\mbox{\scriptsize c}}(L)$ then depend on $L$?

The near-critical order distribution $p_{L,{\cal M}}({\cal M})$ will have a local minimum at ${\cal M} = {\cal M}_{\mbox{\scriptsize c}}(L)$ in accord with the matching requirements. In the absence of $p_{h}^{*}(x)$ and $p_{5}^{*}(x)$ (and any higher order odd terms), this value should coincide with the limiting value ${\cal M}_{\mbox{\scriptsize c}}\: (\equiv {\cal M}_{\mbox{\scriptsize c}}^{\infty})$ --- which merely says that $p_{L,{\cal M}}({\cal M})$ has a local minimum at $x = 0$. However, the antisymmetric corrections, $p_{h}^{*}(x)$ and $p_{5}^{*}(x)$ must shift ${\cal M}_{\mbox{\scriptsize c}}(L)$ away from ${\cal M}_{\mbox{\scriptsize c}}$ so that $p_{L,{\cal M}}({\cal M})$ will have a local minimum at some $x \neq 0$. To find this location, we expand the scaling functions in eq 28 about $x=0$: according to their symmetries one may write
 \begin{eqnarray}
  p_{\cal M}^{*}(x) & = & p_{*}^{0} + p_{*}^{(2)}x^{2} + p_{*}^{(4)}x^{4} + \cdots,  \label{ch5-eq5.30} \\
  p_{t}^{\ast}(x) & = & p_{t0}^{\ast} + p_{t2}^{\ast}x^{2} + p_{t4}^{\ast}x^{4} + \cdots,  \nonumber
 \end{eqnarray}
and similarly for $p_{4}^{\ast}(x)$, while $p_{h}^{\ast}(x)$ and $p_{5}^{\ast}(x)$ generate only odd powers of $x$. At the minimum, the derivative of $p_{L,{\cal M}}({\cal M})$, which takes the form
 \begin{eqnarray}
  p_{L,{\cal M}}^{\prime}({\cal M}) & \propto & 2p_{*}^{(2)}x + 4p_{*}^{(4)}x^{3} + \cdots + 2a_{\cal E}L^{1/\nu}\tilde{t}( p_{t2}^{*}x + 2p_{t4}^{*}x^{3} + \cdots) \nonumber \\
  &  & +\: 2a_{4}L^{-\theta/\nu}(p_{42}^{*}x + 2p_{44}^{*}x^{3} + \cdots) \nonumber \\
  &  & +\: a_{\cal M}\tilde{a}_{h}L^{-(1-\alpha-\beta)/\nu}(p_{h1}^{*}+3p_{h3}^{*}x^{2} + \cdots) \nonumber \\
  &  & +\: a_{5}L^{-\theta_{5}/\nu}(p_{51}^{*} + 3p_{53}^{*}x^{2} + \cdots) + \cdots, 
 \end{eqnarray}
must vanish. (Note that we have used eq 29.) On the other hand, the magnitude matching relation eq 26 yields the further condition
 \begin{eqnarray}
  0 & = & p_{*}^{(2)}x^{2} + p_{*}^{(4)}x^{4} + \cdots + a_{\cal E}L^{1/\nu}\tilde{t}(p_{t0}^{*} + p_{t2}^{*}x^{2} + \cdots) \nonumber \\
  &  & +\: a_{4}L^{-\theta/\nu}(p_{40}^{*} + p_{42}^{*}x^{2} + \cdots) \nonumber \\
  &  & +\: a_{\cal M}\tilde{a}_{h}L^{-(1-\alpha-\beta)/\nu}(p_{h1}^{*}x + p_{h3}^{*}x^{3} + \cdots) \nonumber \\
  &  & +\: a_{5}L^{-\theta_{5}/\nu}(p_{51}^{*}x + p_{53}^{*}x^{3} + \cdots) + \cdots.  \label{ch5-eq5.35}
 \end{eqnarray}
Solving these two conditions simultaneously for $x$ and $\tilde{t}$ yields
 \begin{eqnarray}
  \tilde{t} & = & -(a_{4}p_{40}^{*}/p_{t0}^{\ast}a_{\cal E})/L^{(1+\theta)/\nu} + \cdots,  \label{ch5-eq5.36} \\
  x & = & -\: (a_{\cal M}\tilde{a}_{h}p_{h1}^{*}/2p_{*}^{(2)})/L^{(1-\alpha -\beta)/\nu} + \cdots \nonumber \\
  &  &  -\: (a_{2}p_{31}^{*}/2p_{*}^{(2)})/L^{\theta_{5}/\nu} + \cdots.  \label{ch5-eq5.37}
 \end{eqnarray}
We may then note from the definition of ${\cal R}^{\ast}$ in eq 30, that the $p^{*}$ ratio in the amplitude for $\tilde{t}$ is simply $p_{40}^{*}/p_{t0}^{*} = {\cal R}^{\ast}$. From eqs 23 and 38, we obtain
 \begin{eqnarray}
  \delta {\cal M}(L) & \equiv & {\cal M}_{\mbox{\scriptsize c}}(L) - {\cal M}_{\mbox{\scriptsize c}}^{\infty}, \nonumber \\
  & \approx & A_{1}/L^{(1-\alpha)/\nu} + \cdots + A_{5}/L^{(\beta+\theta_{5})/\nu},  \label{ch5-eq5.39}
 \end{eqnarray}
where $A_{1} = - a_{\cal M}^{2}\tilde{a}_{h}p_{h1}^{*}/2p_{*}^{(2)}$ and $A_{5} = - a_{\cal M}a_{5}p_{51}^{*}/2p_{*}^{(2)}$. In contrast to the arguments of Wilding and M\"{u}ller,$^{6,7}$ we thus find that ${\cal M}_{\mbox{\scriptsize c}}(L)$ {\em does} depend on $L$ with a leading exponent $-(1-\alpha)/\nu$ that, for Ising-type systems, is approximately $-1.41$.$^{23}$ 

To obtain the $L$-dependence of the energy estimator ${\cal E}_{\mbox{\scriptsize c}}(L)$, we first recall that the energy distribution $p_{\cal E}^{*}(y)$ is asymmetric with a maximum at $y=y^{*}\neq 0$: see Fig.\ 2. The matching of $p_{L,{\cal E}}({\cal E})$ to $p_{\cal E}^{*}(y)$ then yields
 \begin{equation}
  \delta {\cal E}(L) \equiv {\cal E}_{\mbox{\scriptsize c}}(L) - {\cal E}_{\mbox{\scriptsize c}}^{\infty} \approx a_{\cal E}y^{*}/L^{(1-\alpha)/\nu}. \label{ch5-eq5.40}
 \end{equation}
Finally, from eq 33 we obtain
 \begin{eqnarray}
  \check{\rho}_{\mbox{\scriptsize c}}(L) - \check{\rho}_{\mbox{\scriptsize c}}^{\infty} & \approx & (A_{1} + l_{1}a_{\cal E}y^{*})/L^{(1-\alpha)/\nu},  \label{ch5-eq5.41} \\
  \check{u}_{\mbox{\scriptsize c}}(L) - \check{u}_{\mbox{\scriptsize c}}^{\infty} & \approx & (a_{\cal E}y^{*} + k_{1}A_{1})/L^{(1-\alpha)/\nu}.  \label{ch5-eq5.42}
 \end{eqnarray}
Notice that the leading exponent $(1-\alpha)/\nu$ agrees with the assertions of refs 6 and 7 so that the dependence of ${\cal M}_{\mbox{\scriptsize c}}(L)$ (contrary to the related claims$^{6,7}$) does not actually disturb the anticipated asymptotic behavior.

This completes our analysis of the BW approach when pressure mixing may be neglected.

\ssection{\hspace{-.27in}.\hspace{0.1in}Inclusion of Pressure Mixing}
\indent

As discussed in the Introduction, it is important in studying asymmetric fluid criticality to understand and, if possible, to allow for the effects of pressure mixing in the BW approach. In ref 37 a scaling expression for the bulk canonical free energy density, $f(\rho,T)$, that incorporates pressure mixing is derived. On this basis a finite-size scaling form for the singular part of the canonical free energy density $f(\rho,T;L)$ was advanced. A crucial feature is that pressure mixing introduces an extra correction term that vanishes as $j_{2}/L^{\beta/\nu}$ when $L\rightarrow \infty$. This contribution turns out to be antisymmetric in the ordering operator $\delta {\cal M}$. On noticing that $\ln [p_{L,{\cal M}}({\cal M})]$ becomes closely related to $f(\rho,T)$ when $L\rightarrow\infty$, we see that the scaling ansatz for the order distribution $p_{L,{\cal M}}({\cal M})$ postulated by BW$^{2,3}$ should better be modified to read, in expanded form [compare with eq 28]
 \begin{eqnarray}
  p_{L,M}({\cal M}) & \approx & {\cal N}_{\cal M}\left[ p_{\cal M}^{*}(x) + a_{\cal E}L^{1/\nu}\tilde{t}p_{t}^{*}(x) + j_{2}a_{j}L^{-\beta/\nu}p_{j}^{*}(x) + a_{4}L^{-\theta/\nu}p_{4}^{\ast}(x) \right. \nonumber \\
  &  & \hspace{0.65in} +\: a_{\cal M}L^{\Delta/\nu}\tilde{h} p_{h}^{*}(x) + a_{5}L^{-\theta_{5}/\nu}p_{5}^{*}(x) + \cdots],  \label{ch5-eq5.43}
 \end{eqnarray}
where the new, i.e., the third term entails the scaling function $p_{j}^{*}(x)$ which should be universal and antisymmetric in $x$ while the amplitude, $a_{j}$, is nonuniversal. In fact, this new contribution dominates all subsequent corrections when the exponents take Ising or similar values (where we appeal to eq 29 to see that $L^{\Delta/\nu}\tilde{h}\propto L^{-(1-\alpha-\beta)/\nu}$).

The presence of this pressure mixing term evidently raises a further question about the validity of the BW method. When $j_{2}$ is small, one may well still obtain good matching of the observed distribution $p_{L,{\cal M}}({\cal M})$ to the symmetric fixed point function $p_{\cal M}^{*}(x)$ even for relatively small system sizes. The 2D Lennard-Jones fluid may represent such a case, since Wilding$^{7}$ observed that $p_{L,{\cal M}}({\cal M})$ could be well symmetrized and readily matched to $p_{\cal M}^{*}(x)$. However, if $j_{2}$ is sufficiently large, one should not ignore its contribution: then symmetrization of $p_{L,{\cal M}}({\cal M})$ should be feasible only in an approximate way even for relatively large $L$. Indeed, Caillol, Levesque and Weis$^{12}$ performed Monte Carlo simulations on equicharged hard-spheres (i.e., the RPM electrolyte) and observed that their data for $p_{L,{\cal M}}({\cal M})$ could be matched to the Ising distribution $p_{\cal M}^{*}(x)$ only for large $L$; for small $L$ they were unable to symmetrize via the BW procedure. They attributed this failure to poor data sampling in the low density region of their smaller systems; but it would seem that significant pressure mixing in the model$^{35}$ could well be the primary cause of the observed asymmetry although the antisymmetric contribution due to $\tilde{h}$, which varies like $L^{-(1-\alpha-\beta)/\nu}$ [see eq 29], may also be a factor. Further simulations to resolve these possibilities would, thus, be interesting. Nevertheless, $p_{L,{\cal M}}^{\mbox{\scriptsize c}}({\cal M})$ should always asymptotically approach $p_{\cal M}^{*}(x)$ when $L\rightarrow \infty$. In practice therefore, one may still be able to match $p_{L,{\cal M}}({\cal M})$ to $p_{\cal M}^{*}(x)$ within tolerable precision for large enough $L$ and thence derive best-fit estimators via the BW recipe. How will these then depend on the system size $L$?

Before addressing this question, we will revisit the matching conditions described in section III.1. As demonstrated by eqs 9 and 10, we must expect the critical ordering and energy densities, $\delta{\cal M}$ and $\delta{\cal E}$, to actually be {\em nonlinear} combinations of the density and energy fluctuations, $\delta \check{\rho}$ and $\delta\check{u}$. Near enough to the critical point, when the typical deviations are small, however, linear BW relations, following eqs 13 and 14, should still become valid. But we expect such linear relations to be inadequate further from criticality.

Even if one can reach regimes where the linear relations are valid, however, the matching procedure should be more complicated, in order to accommodate the two extra unknown parameters, $j_{1}$ and $j_{2}$. Here we propose an approach which, as far as possible, adopts the steps presented previously in section III.1: {\bf (i)} First suppose $j_{1}=j_{2}=0$ and proceed through steps {\bf (a)}-{\bf (c)} in order to match the data for $p_{L,{\cal M}}({\cal M})$ as well as feasible to $p_{\cal M}^{*}(x)$ and so obtain the first round of estimators $\Tc^{(1)}(L)$, $\mu_{\mbox{\scriptsize c}}^{(1)}(L)$, $l_{1}^{(1)}(L)$, $a_{\cal M}^{(1)}(L)$, and ${\cal M}_{\mbox{\scriptsize c}}^{(1)}(L)$. The remaining parameters, ${\cal E}_{\mbox{\scriptsize c}}^{(1)}(L)$, $k_{1}^{(1)}(L)$ and $a_{\cal E}^{(1)}(L)$ can be obtained similarly by matching the energy operator distribution $p_{L,{\cal E}}({\cal E})$ to the fixed point distribution $p_{\cal E}^{*}(y)$ as explained before. We should also recall the relations  $l_{0}^{(1)}(L) = \check{\rho}_{\mbox{\scriptsize c}}^{(1)}(L)$ and $k_{0}^{(1)}(L) = -\check{u}_{\mbox{\scriptsize c}}^{(1)}(L)$. At this stage, however, one may still observe differences between the mixed data distributions and the fixed point distributions, especially further from criticality. {\bf (ii)} Knowing via eq 43 that $j_{2}$ relates to asymmetry in $p_{L,{\cal M}}({\cal M})$, we now suppose, as a tentative approximation, that the fluctuating critical densities, $\delta{\cal M}$ and $\delta{\cal E}$, are related to the observable fluctuations, $\delta\check{\rho}$ and $\delta \check{u}$, via {\em nonlinear relations} that parallel eqs 9 and 10 (i.e., obtained by removing the expectation brackets and replacing the $\Delta$'s by $\delta$'s). Then, on first retaining the setting $j_{1}=0$, one can attempt to adjust $j_{2}$ to obtain a value, say $j_{2}^{(1)}(L)$, that provides a better match of $p_{L,{\cal M}}({\cal M})$ to $p_{\cal M}^{*}(x)$. Next {\bf (iii)} one may adjust $j_{1}$ in the nonlinear relations to improve the matching of $p_{L,{\cal E}}({\cal E})$ to $p_{\cal E}^{*}(y)$: this should yield a value $j_{1}^{(1)}(L)$. {\bf (iv)} Now one can set $j_{1}=j_{1}^{(1)}(L)$ and $j_{2}=j_{2}^{(1)}(L)$ and recalculate the order distribution $p_{L,{\cal M}}({\cal M})$: one is likely to observe some new discrepancies near the local minimum and the two maxima. Accordingly, one can return to step {\bf (i)} but now with fixed $j_{1}=j_{1}^{(1)}(L)$ and $j_{2}=j_{2}^{(1)}(L)$, and iterate the procedure to find a new set of parameters, say $\Tc^{(2)}(L)$, $\mu_{\mbox{\scriptsize c}}^{(2)}(L)$, etc. On repeating these steps, one should be led to stable values for all the parameters. Nevertheless, as a practical matter one may reasonably question the robustness of this procedure (which we have {\em not} ourselves attempted to implement).

To obtain an assessment of the effect of pressure mixing on the convergence of the BW procedure, however, it suffices to suppose that we have achieved a good matching for the distribution functions $p_{L,{\cal M}}({\cal M})$ and $p_{L,{\cal E}}({\cal E})$ and have in hand a satisfactory fit of the critical parameters. To understand the $L$-dependence, we first expand the new universal function $p_{j}^{*}(x)$ in eq 43 as
 \begin{equation}
  p_{j}^{*}(x) = p_{j1}^{*}x + p_{j3}^{*}x^{3} + \cdots,  \label{ch5-eq5.44}
 \end{equation}
while the other functions appearing in eq 43 may be expanded just as in Sec.\ III.3: see eq 34. The local minimum of $p_{L,{\cal M}}({\cal M})$ must then satisfy
 \begin{eqnarray}
  0 & = & 2p_{*}^{(2)}x + 4p_{*}^{(4)}x^{3} + \cdots + 2a_{\cal E}L^{1/\nu}\tilde{t}(p_{t2}^{*}x + 2p_{t4}^{*}x^{3} + \cdots) \nonumber \\
  &  & +\: j_{2}a_{j}L^{-\beta/\nu}(p_{j1}^{*} + 3p_{j3}^{*}x^{2} + \cdots) +2a_{4}L^{-\theta/\nu}(p_{42}^{*}x + 2p_{44}^{*}x^{3} + \cdots) \nonumber \\
  &  & +\: a_{\cal M}\tilde{a}_{h}L^{-(1-\alpha-\beta)/\nu}(p_{h1}^{*} + 3p_{h3}^{*}x^{2} + \cdots) + a_{5}L^{-\theta_{5}/\nu}(p_{51}^{*} + 3p_{53}^{*}x^{2} + \cdots) + \cdots.  \label{ch5-eq5.45}
 \end{eqnarray}
while the matching condition, $p_{L,{\cal M}}({\cal M}_{\mbox{\scriptsize c}}) = p_{\mbox{\scriptsize min}}^{\ast\cal M}$, yields a further condition corresponding to eq 36, namely,
 \begin{eqnarray} 
  0 & = & p_{*}^{(2)}x^{2} + p_{*}^{(4)}x^{4} + \cdots + a_{\cal E}L^{1/\nu}\tilde{t}(p_{t0}^{*} + p_{t2}^{*}x^{2} + \cdots) \nonumber \\
  &  & +\: j_{2}a_{j}L^{-\beta/\nu}(p_{j1}^{*}x + p_{j3}^{*}x^{3} + \cdots) + a_{4}L^{-\theta/\nu}(p_{40}^{*} + p_{42}^{*}x^{2} + \cdots) \nonumber \\
  &  & +\: a_{\cal M}\tilde{a}_{h}L^{-(1-\alpha-\beta)/\nu}(p_{h1}^{*}x + p_{h3}^{*}x^{3} + \cdots) + a_{5}L^{-\theta_{5}/\nu}(p_{51}^{*}x + p_{53}^{*}x^{3} + \cdots) + \cdots.  \label{ch5-eq5.46}
 \end{eqnarray}
Solving these two equations simultaneously for $x$ and $\tilde{t}$ yields
 \begin{equation}
  2p_{\ast}^{(2)}x = -\: j_{2}a_{j}p_{j1}^{*}/L^{\beta/\nu} - a_{\cal M}\tilde{a}_{h}p_{h1}^{*}/L^{(1-\alpha-\beta)/\nu} - 2j_{2}a_{j}a_{4}p_{j1}^{*}p_{42}^{*}/L^{(\beta+\theta)/\nu} + \cdots,  \label{ch5-eq5.47}
 \end{equation}
while, as regards the leading term, $\tilde{t}$ is still given by eq 37. Note that the $L^{-\theta_{5}/\nu}$ term, that originally appeared in eq 38 (when $j_{1}=j_{2}=0$) still arises but only as a higher order correction not displayed here. [As previously, we assume tacitly that the exponent values lie within the normal range of $O(n)$ fixed points.]

To derive the modified form for $\Tc(L)$, we first rewrite eq 2, the definition for $\tilde{p}$, in the form
 \begin{equation}
  \check{p} = k_{0}t + l_{0}\check{\mu} + \tilde{p} + \cdots.  \label{ch5-eq5.48}
 \end{equation}
On using eq 3, the definition of $\tilde{h}$, similarly, the result in eq 29 then yields
 \begin{equation}
  \check{\mu} = \frac{k_{1}+j_{2}k_{0}}{1-j_{2}l_{0}}t + \frac{1}{1-j_{2}l_{0}}\tilde{p} + \frac{\tilde{a}_{h}}{1-j_{2}l_{0}}L^{-(1-\alpha+\gamma)/\nu} + \cdots,  \label{ch5-eq5.49}
 \end{equation}
which, on substitution, gives the leading order result
 \begin{equation}
  \check{p} = \left(k_{0} + l_{0}\frac{k_{1}+j_{2}k_{0}}{1-j_{2}l_{0}}\right) t + \cdots,  \label{ch5-eq5.50}
 \end{equation}
from which we have dropped the terms varying as $\tilde{p} \propto L^{-(2-\alpha)/\nu}$ (see ref 37) and $L^{-(1-\alpha+\gamma)/\nu}$ which  enter only as higher order corrections. We may now substitute these results into eq 37 and use eq 4 for $\tilde{t}$ to find
 \begin{equation}
  t_{\mbox{\scriptsize c}}(L) = \frac{\Tc(L) - \Tc^{\infty}}{\Tc^{\infty}} = -\frac{a_{1}{\cal R}^{\ast}}{\tau a_{\cal E}}/L^{(1+\theta)/\nu} + \cdots,  \label{ch5-eq5.51}
 \end{equation}
where the mixing coefficient combination is
 \begin{equation}
  \tau = 1-j_{1}k_{0} - (l_{1}+j_{1}l_{0})(k_{1}+j_{2}k_{0})/(1-j_{2}l_{0}), \label{ch5-eq5.52}
 \end{equation}
while ${\cal R}^{\ast}=p_{40}^{\ast}/p_{t0}^{\ast}$ as in eq 38. The chemical potential may be obtained by substitution in eq 49 which yields
 \begin{equation}
  \mu_{\mbox{\scriptsize c}}(L) - \mu_{\mbox{\scriptsize c}}^{\infty} \approx \frac{(k_{1}+j_{2}k_{0})a_{1}{\cal R}^{\ast}k_{\mbox{\scriptsize B}}\Tc}{(1-j_{2}l_{0})\tau a_{\cal E}}/L^{(1+\theta)/\nu}.  \label{ch5-eq5.53}
 \end{equation}
Note that pressure mixing does not alter the leading exponent but does change the amplitude.

Finally, the critical order estimator ${\cal M}_{\mbox{\scriptsize c}}(L)$ can be obtained from eq 47 which leads to the replacement of eq 39 by
 \begin{equation}
  \delta {\cal M} \approx - j_{2}A_{2}/L^{2\beta/\nu} + A_{1}/L^{(1-\alpha)/\nu} - j_{2}A_{4}/L^{(2\beta+\theta)/\nu},  \label{ch5-eq5.54}
 \end{equation}
where $A_{2}=a_{\cal M}a_{j}p_{j1}^{\ast}/2p_{\ast}^{(2)}$ and $A_{4}=2a_{4}p_{42}^{\ast}A_{2}$, while $A_{1}$ is given after eq 39. On the other hand, to leading order, eq 40 remains valid for the energy estimator, ${\cal E}_{\mbox{\scriptsize c}}(L)$. To complete the calculation we now invert eqs 9 and 10 to obtain, up to linear order,
 \begin{eqnarray}
  \Delta\check{\rho} & = &  (1-j_{2}l_{0})\delta {\cal M} + (l_{1}+j_{1}l_{0})\delta {\cal E},  \label{ch5-eq5.55} \\
  \Delta\check{u} & = & (1-j_{1}k_{0})\delta {\cal E} + (k_{1}+j_{2}k_{0})\delta {\cal M}.  \label{ch5-eq5.56}
 \end{eqnarray}
Appealing to eqs 40 and 54 then yields our main conclusions, namely,
 \begin{eqnarray}
  \check{\rho}_{\mbox{\scriptsize c}}(L) - \check{\rho}_{\mbox{\scriptsize c}}(\infty) & = & B_{\rho}L^{-2\beta/\nu} + A_{\rho}L^{-(1-\alpha)/\nu} + B_{4\rho}L^{-(2\beta+ \theta)/\nu} + \cdots,  \label{ch5-eq5.57} \\
  \check{u}_{\mbox{\scriptsize c}}(L) - \check{u}_{\mbox{\scriptsize c}}(\infty) & = & B_{u}L^{-2\beta/\nu} + A_{u}L^{-(1-\alpha)/\nu} + B_{4u}L^{-(2\beta+\theta)/\nu} + \cdots,  \label{ch5-eq5.58}
 \end{eqnarray}
where the coefficients are
 \begin{eqnarray}
  B_{\rho} & = & - j_{2}(1-j_{2}l_{0})A_{2}, \hspace{0.15in} A_{\rho} = (1-j_{2}l_{0})A_{1} + (l_{1}+j_{1}l_{0})a_{\cal E}y^{*},  \label{ch5-eq5.59d} \\
  B_{u} & = & - j_{2}(k_{1}+j_{2}k_{0})A_{2},  \hspace{0.15in} A_{u} = (1-j_{1}k_{0})a_{\cal E}y^{*} + (k_{1}+j_{2}k_{0})A_{1}, \label{ch5-eq5.59c} \\
  B_{4\rho} & = & 2a_{1}p_{22}^{*}B_{\rho} \hspace{0.33in} \mbox{and} \hspace{0.3in}  B_{4u} = 2a_{1}p_{22}^{*} B_{u}.   \label{ch5-eq5.59b}
 \end{eqnarray}
When the pressure-mixing coefficient $j_{2}$ vanishes, the leading $L^{-2\beta/\nu}$ terms drop out but the $L^{-(1-\alpha)/\nu}$ terms remain; in that case one regains the original BW exponents for $\rhoc(L)$ and $u_{\mbox{\scriptsize c}}(L)$ although the amplitudes now depend on $j_{1}$. Evidently pressure mixing may significantly slow the rate of convergence in estimating $\rhoc$ and $u_{\mbox{\scriptsize c}}$. In practice it may be more significant that the exponents $2\beta/\nu\simeq 1.03$, $(1-\alpha)/\nu\simeq 1.41$, and $(2\beta+\theta)/\nu\simeq 1.86$ (using Ising values) are fairly close in magnitude so that if the successive terms are of different sign and thus compete, reliable extrapolation may be seriously hampered. This could be the cause of the misleading BW error estimate for $\rhoc$ seen in Table 1 for the RPM.

\ssection{\hspace{-.27in}.\hspace{0.1in}Conclusions}
\indent

In summary we have examined critically the Bruce-Wilding extrapolation method that, in the past, has been widely applied to various model systems since it usually provides a straightforward and apparently reliable way of estimating critical parameters from finite-size data. We first analyzed in some detail the original BW method, that neglects pressure mixing in the scaling fields. Effective critical parameter estimators, $\Tc(L)$, $\rhoc(L)$, etc., can be obtained by matching the numerically measured distribution functions $p_{L,{\cal M}}({\cal M})$ and $p_{L,{\cal E}}({\cal E})$ to the fixed point functions $p_{\cal M}^{*}(x)$ and $p_{\cal E}^{*}(y)$, respectively. We provided a precise specification for implementing the matching procedure that generates satisfactory fits and yields explicit values for the critical parameter estimators that can be investigated analytically. The finite-size behavior of the estimators, $\Tc(L)$, $\rhoc(L)$, etc., was then derived. The principal BW claims, namely, that $\Delta\Tc(L)$ and $\Delta\mu_{\mbox{\scriptsize c}}(L)$ decay like $1/L^{(1+\theta)/\nu}$ when $L\rightarrow\infty$ while $\Delta\rhoc(L)$ and $\Delta u_{\mbox{\scriptsize c}}(L)$ vanish as  $1/L^{(1-\alpha)/\nu}$ were confirmed with explicit expressions for the amplitudes: see eqs 31, 32, 41, and 42.

When pressure mixing is allowed for, however, the ordering operator distribution, $p_{L,{\cal M}}({\cal M})$, contains extra {\em antisymmetric} corrections that vanish only as $1/L^{\beta/\nu}$; as a result these dominate all other corrections to scaling. Consequently, the matching of $p_{L,{\cal M}}({\cal M})$ to the symmetric fixed point function $p_{\cal M}^{*}(x)$ can be achieved only in an approximate way for finite $L$ if one follows the BW recipe. An extension of the BW procedure that makes some allowance for pressure mixing was proposed but has not been tested. Nevertheless, by assuming that an acceptable matching can be realized, we demonstrated that pressure mixing does not alter the leading $1/L^{(1+\theta)/\nu}$ term in the effective critical temperature, $\Tc(L)$: see eq 51 (although the amplitude is changed). On the other hand, the effective critical (number) density, $\rhoc(L)$, and energy density, $u_{\mbox{\scriptsize c}}(L)$, contain new, $1/L^{2\beta/\nu}$, terms with amplitudes proportional to the pressure-mixing coefficient $j_{2}$. Furthermore, these terms asymptotically dominate those previously identified: see eqs 57 and 58.

In conclusion, the Bruce-Wilding method may still be regarded as a useful practical guide to the extrapolation of finite-size simulation data for systems that do not deviate far from symmetry and that may with confidence be expected to fall within the Ising critical universality class: see Table 1 for some indications of its precision and numerical short comings. In other cases, however, the first issue of concern is that one lacks numerically reliable universal fixed-point distributions for the order parameter and energy that are essential for the method. Indeed, as outlined in the Introduction, Camp and Patey$^{17}$ implemented the BW method in a study of liquid-gas criticality in model fluids with algebraically decaying attractive pair interactions, and encountered just this problem! Nevertheless, even if the required universal distributions were accurately known, one could not reasonably expect to benefit from the fortunate but apparently quite accidental cancelation of the thermal and leading even-correction scaling functions which have greatly assisted practical BW calculations for Ising systems: see the discussion in association with eq 30 (where the ratio ${\cal R}^{\ast}$ was introduced).

Beyond that, even when Ising criticality may be confidently assumed, the presence of {\em both} odd-order correction terms {\em and} significant pressure mixing, must be expected if, as in the case of the hard-sphere ionic fluid models,$\,^{38}$ the observed asymmetries are not small. In such cases, even when the BW matching recipe can be implemented satisfactorily, the results will very likely be distorted (relative to the more symmetric cases); consequently, the subsequent extrapolations must be regarded with increased caution and assessed as less reliable: see Table 1.

While we have sketched one iterative BW-type method that could allow for pressure mixing, one might consider further elaborations of the BW approach --- for example, by directly examining and ``tuning out'' the scaled {\em cross correlations} of the number density, $\rho$, and the configurational energy density, $u$. However, in the light of the recently developed bias-free procedures, involving the $Q$-loci and related estimators$^{34,35,38}$ (which seem to work rather reliably for at least some highly asymmetric systems and do {\em not} require detailed prior knowledge), attempts to further extend the BW approach do not seem warranted at present.

\pagebreak
\hspace{.2in}{\large \bf Acknowledgments}
\indent

 We are grateful to Erik Luijten for valuable comments on an earlier account of the work reported here. The support of the National Science Foundation (through Grant No. CHE 03-01101) has been appreciated.

\pagebreak
\hspace{-.2in}{\large \bf References and Notes}
\begin{itemize}
 \item[(1)] See, e.g.: (a) Binder, K. in {\em Computational Methods in Field Theory}, edited by Gausterer, H.\ and Lang, C.\ B.\ (Springer, Berlin, 1992) pp. 59-125; (b) Privman, V. editor, {\em Finite-Size Scaling and Numerical Simulation of Statistical Systems} (World Scientific, Singapore, 1990).
 \item[(2)] Bruce, A.\ D.; Wilding, N.\ B.\ {\em Phys. Rev. Lett.} {\bf 1992}, {\em 68}, 193. 
 \item[(3)] Wilding, N.\ B.; Bruce, A.\ D.\ {\em J.\ Phys.: Condens.\ Matter} {\bf 1992}, {\em 4}, 3087.
 \item[(4)] Wilding, N.\ B.\ {\em Z.\ Phys.\ B} {\bf 1993}, {\em 93}, 119.
 \item[(5)] Hilfer, R.; Widling, N.\ B.\ {\em J.\ Phys.\ A: Math.\ Gen.} {\bf 1995}, {\em 28}, L281.
 \item[(6)] Wilding, N.\ B.; M\"{u}ller, M.\ {\em J.\ Chem.\ Phys.} {\bf 1995}, {\em 102}, 2562.
 \item[(7)] Wilding, N.\ B.\ {\em Phys.\ Rev.\ E} {\bf 1995}, {\em 52}, 602.
 \item[(8)] M\"{u}ller, M.; Wilding, N.\ B.\ {\em Phys.\ Rev.\ E} {\bf 1995}, {\em 51}, 2079.
 \item[(9)] Wilding, N.\ B.\ {\em J.\ Phys.: Condens.\ Matter} {\bf 1996}, {\em 8}, 9637.
 \item[(10)] Wilding, N.\ B.; Nielaba, P.\ {\em Phys.\ Rev.\ E} {\bf 1996}, {\em 53}, 926.
 \item[(11)] Wilding, N.\ B.\ {\em Phys.\ Rev.\ E} {\bf 1997}, {\em 55}, 6624.
 \item[(12)] Caillol, J.\ M.; Levesque, D.; Weis, J.\ J.\ {\em J.\ Chem.\ Phys.} {\bf 1997}, {\em 107}, 1565.
 \item[(13)] Caillol, J.\ M.\ {\em J.\ Chem.\ Phys.} {\bf 1998}, {\em 109}, 4885.
 \item[(14)] Orkoulas, G.; Panagiotopoulos, A.\ Z.\ {\em J.\ Chem.\ Phys.} {\bf 1998}, {\em 110}, 1581.
 \item[(15)] Romero-Enrique, J.\ M.; Orkoulas, G.; Panagiotopoulos, A.\ Z.; Fisher, M.\ E.\ {\em Phys.\ Rev.\ Lett.} {\bf 2000}, {\em 85}, 4558.
 \item[(16)] Yan, Q.\ L.; de Pablo, J.\ J. {\em J.\ Chem.\ Phys.} {\bf 2001}, {\em 114}, 1727; {\em Phys.\ Rev.\ Lett.} {\bf 2001}, {\em 86}, 2054.
 \item[(17)] Camp, P.\ J.; Patey, G.\ N.\ {\em J.\ Chem.\ Phys.} {\bf 2001}, {\em 114}, 399.
 \item[(18)] Panagiotopoulos, A.\ Z.\ {\em J.\ Chem.\ Phys.} {\bf 2002}, {\em 116}, 3007.
 \item[(19)] Panagiotopoulos, A.\ Z.; Fisher, M.\ E.\ {\em Phys.\ Rev.\ Lett.} {\bf 2002}, {\em 88}, 045701.
 \item[(20)] Moghaddam, S.; Panagiotopoulos, A.\ Z.\ {\em J.\ Chem.\ Phys.} {\bf 2003}, {\em 118}, 7556.
 \item[(21)] Mermin, N.\ D.\ {\em Phys.\ Rev.\ Lett.} {\bf 1971}, {\em 26}, 169.
 \item[(22)] Rehr, J.\ J.; Mermin, N.\ D.\ {\em Phys. Rev. A} {\bf 1973}, {\em 8}, 472.
 \item[(23)] Kim, Y.\ C.; Fisher, M.\ E.; Barbosa, M.\ C.\ {\em J.\ Chem.\ Phys.} {\bf 2001}, {\em 115}, 933. This reference may be consulted for definitions and values of the standard critical exponents: see also refs 29, 31, and 37 below.
 \item[(24)] Nicolaides, D.; Bruce, A.\ D.\ {\em J.\ Phys.\ A: Math.\ Gen.} {\bf 1988}, {\em 21}, 233.
 \item[(25)] Tsypin, M.\ M.; Bl\"{o}te, H.\ W.\ J.\ {\em Phys.\ Rev.\ E} {\bf 2000}, {\em 62}, 73.
 \item[(26)] Fisher, M.\ E.; Ma, S.; Nickel, B.\ G.\ {\em Phys.\ Rev.\ Lett.} {\bf 1972}, {\em 29}, 917.
 \item[(27)] Sak, J.\ {\em Phys.\ Rev.\ B} {\bf 1973}, {\em 8}, 281.
 \item[(28)] Hilfer, R.\ {\em Z.\ Phys.\ B: Condens.\ Matter}, {\bf 1994}, {\em 96}, 63.
 \item[(29)] Weing\"{a}rtner, H.; Schr\"{o}er, W.\ {\em Adv.\ Chem.\ Phys.} {\bf 2001}, {\em 116}, 1: see, especially, Sec.\ III.B.
 \item[(30)] Fisher, M.\ E.; Orkoulas, G.\ {\em Phys. Rev. Lett.} {\bf 2000}, {\em 85}, 696.
 \item[(31)] Kim, Y.\ C.; Fisher, M.\ E.; Orkoulas, G.\ {\em Phys.\ Rev.\ E} {\bf 2003}, {\em 67}, 061506.
 \item[(32)] Yang, C.\ N.; Yang, C.\ P.\ {\em Phys.\ Rev.\ Lett.} {\bf 1964}, {\em 13}, 303.
 \item[(33)] Orkoulas, G.; Fisher, M.\ E.; \"{U}st\"{u}n, C.\ {\em J.\ Chem.\ Phys.} {\bf 2000}, {\em 113}, 7530.
 \item[(34)] Orkoulas, G.; Fisher, M.\ E.; Panagiotopoulos, A.\ Z.\ {\em Phys.\ Rev.\ E} {\bf 2001}, {\em 63}, 051507.
 \item[(35)] Kim, Y.\ C.; Fisher, M.\ E.; Luijten, E.\ {\em Phys.\ Rev.\ Lett.} {\bf 2003}, {\em 91}, 065701.
 \item[(36)] Kostrowicka Wyczalkowska, A.; Anisimov, M.\ A.; Sengers, J.\ V.; Kim, Y.\ C.\ {\em J.\ Chem.\ Phys.} {\bf 2002}, {\em 116}, 4202.
 \item[(37)] Kim, Y.\ C.; Fisher, M.\ E.\ {\em Phys.\ Rev.\ E} {\bf 2003}, [in press] arXiv:cond-mat/0306331 (12 Jun 2003).
 \item[(38)] Luijten, E.; Fisher, M.\ E.; Panagiotopoulos, A.\ Z.\ {\em Phys.\ Rev.\ Lett.} {\bf 2002}, {\em 88}, 185701.
 \item[(39)] For convenience the notation of ref 37 has been changed here from $(x_{L},y_{L},y_{Lk};D_{L},U_{L},U_{Lk})$ to $(w_{L},z_{L},z_{Lk};a_{\cal E},a_{\cal M},a_{k})$.
 \item[(40)] For hyperscaling in the context of finite-size systems see: Privman, V.; Fisher, M.\ E.\ {\em Phys.\ Rev.\ B} {\bf 1984}, {\em 30}, 322.
 \item[(41)] We neglect here, with confidence since we are concerned only with finite systems with well behaved underlying Boltzmann distributions, those pathological situations in which knowledge of all the moments may {\em not} be adequate to reconstruct the full distribution: see, e.g. {\em Pad\'{e} Approximants; Second edition}, Baker, G.\ A.\ Jr.; Graves-Morris, P.\ (Cambridge University Press, 1996) pp.\ 213-264.
\end{itemize}

\begin{table}
\caption{Estimates for the reduced critical parameters of the hard-core square-well (HCSW) fluid and of the restricted primitive model (RPM) electrolyte via the Bruce-Wilding (BW) method$^{14,18}$ and via bias-free methods.$^{34,38}$ Parentheses denote stated uncertainties in the last decimal place quoted.}
\vspace{0.2in}
\begin{center}
\begin{tabular}{|c|c|c|c|c|}  \hline
  &  \multicolumn{2}{c|}{HCSW}  &  \multicolumn{2}{c|}{RPM} \\  \cline{2-5}
  &  $\Tc^{\ast}$ & $\rhoc^{\ast}$ & $\Tc^{\ast}$ & $\rhoc^{\ast}$ \\ \hline
  BW  & 1.2180(2) & 0.310(1) & 0.05065(2) & 0.084(1) \\ \hline
  Bias-free & 1.2179(3) & 0.3067(4) & 0.05069(2) & 0.0790(25) \\ \hline
\end{tabular}
\end{center}
\vspace{5in}
\end{table}

\pagebreak
\newpage
\begin{figure}
\vspace{-2.5in}
\hspace{2.5in} Figure Captions

\vspace{.3in}
\caption{The universal critical-point order-parameter distribution, $p^{\ast}_{\cal M}(x)$, as a function of the scaled order parameter $x = a_{\cal M}^{-1}L^{\beta/\nu}\delta {\cal M}$, for the Ising universality class, as calculated by Wilding and M\"{u}ller$^{8}$ via Monte Carlo simulations of the $(d$$\,=\,$$3)$-dimensional Ising model. The nonuniversal amplitude $a_{\cal M}$ has been chosen so that the distribution has unit variance; the peaks of height $p^{\ast{\cal M}}_{\mbox{\scriptsize max}}\simeq 0.4267$ are then located at $x = \pm x^{\ast}$ with $x^{\ast}\simeq 1.1801$, while the height of the minimum at $x=0$ is $p_{\mbox{\scriptsize min}}^{\ast\cal M}\simeq 0.1904$.}

\vspace{.3in}
\caption{The universal critical-point energy distribution, $p_{\cal E}^{*}(y)$, for the Ising universality class as a function of the scaled energy $y=a_{\cal E}^{-1}L^{(1-\alpha)/\nu}\delta {\cal E}$, as calculated by Wilding and M\"{u}ller,$^{8}$ selecting the nonuniversal amplitude $a_{\cal E}$ so that the distribution has unit variance. The single peak of height $p^{\ast{\cal E}}_{\mbox{\scriptsize max}}\simeq 0.3981$ occurs at $y^{\ast}\simeq -0.3966$.}

\end{figure}

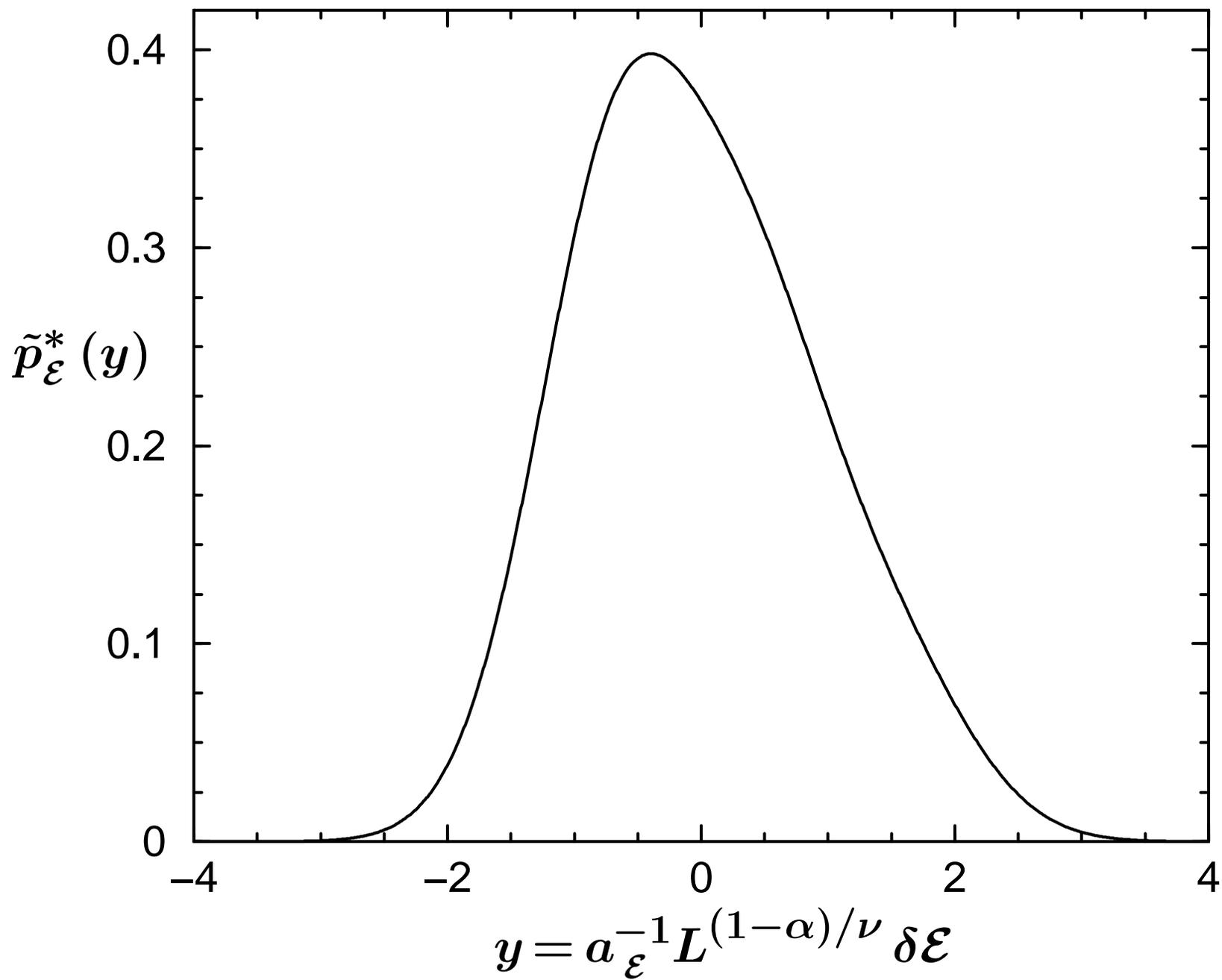
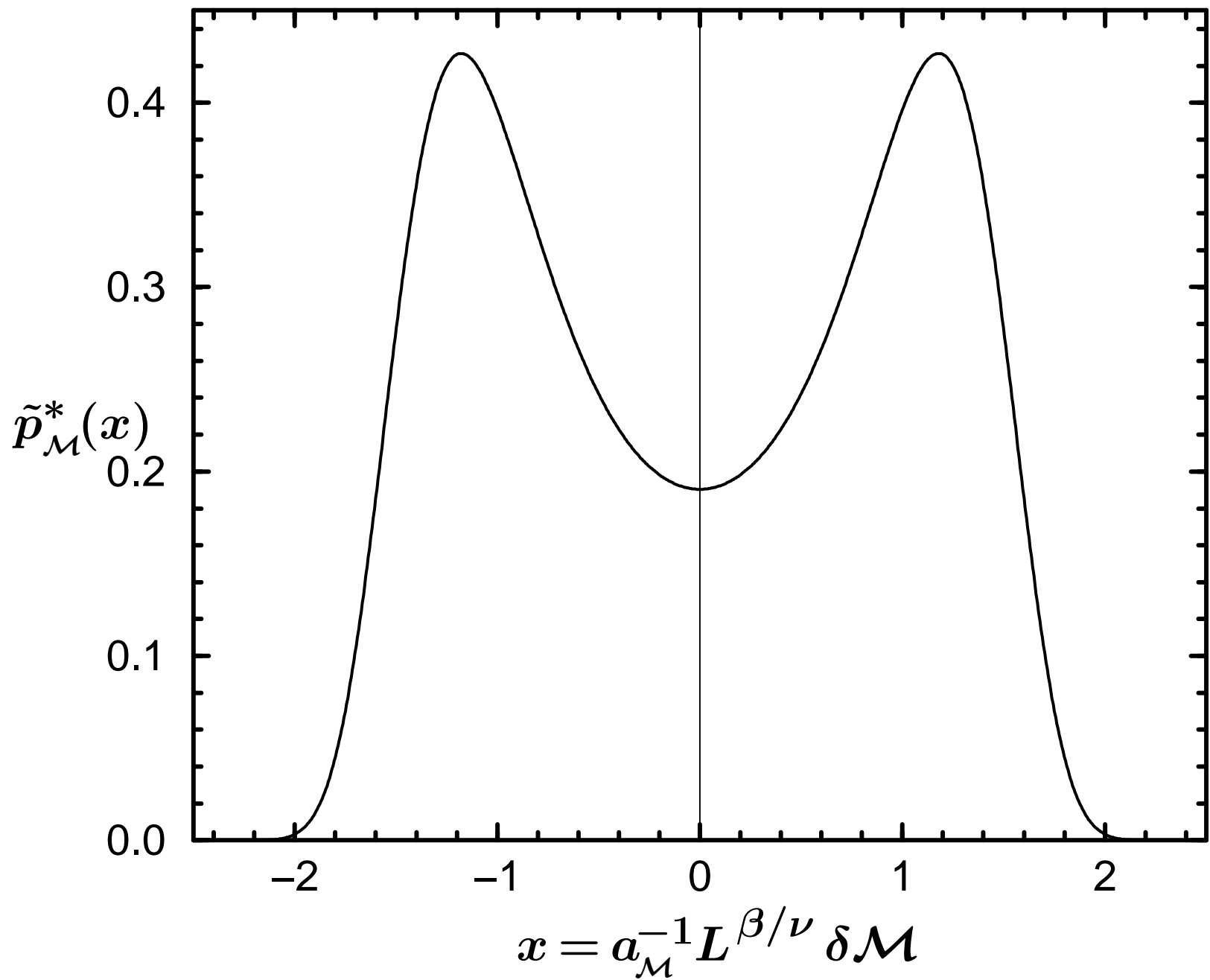
\end{document}